\documentclass[preprint,numerical,onecolumn,superscriptaddress]{revtex4-1}
\usepackage{amssymb, amsmath,framed}
\usepackage{mathtools}
\usepackage{amsfonts,mathrsfs}
\usepackage{subfig}
\usepackage{bm}
\usepackage{tikz}
\usepackage[colorlinks=true,linkcolor=blue]{hyperref}
\expandafter\ifx\csname package@font\endcsname\relax\else
\expandafter\expandafter
\expandafter\usepackage
\expandafter\expandafter
\expandafter{\csname package@font\endcsname}
\fi
\def\bq{\begin{equation}}
\def\eq{\end{equation}}
\def\bqy{\begin{eqnarray}}
\def\eqy{\end{eqnarray}}
\def\bs{\boldsymbol}

\def\p{\partial}

\def\rh{\rho}

\def\p{\partial}

\def\calb{\mathcal{B}}

\begin{document}
\title{On the structure and statistical theory of turbulence of extended magnetohydrodynamics}
\author{George Miloshevich}
\email{gmilosh@physics.utexas.edu}
\affiliation{Department of Physics and Institute for Fusion Studies, The University of Texas at Austin, Austin, TX 78712, USA}
\author{Manasvi Lingam}
\email{manasvi@seas.harvard.edu}
\affiliation{Harvard John A. Paulson School of Engineering and Applied Sciences, Harvard University, Cambridge, MA 02138, USA}
\affiliation{Institute for Theory and Computation, Harvard-Smithsonian Center for Astrophysics, Cambridge, MA 02138, USA}
\affiliation{Department of Astrophysical Sciences, Princeton University, Princeton, NJ 08544, USA}
\author{Philip J. Morrison}
\email{morrison@physics.utexas.edu}
\affiliation{Department of Physics and Institute for Fusion Studies, The University of Texas at Austin, Austin, TX 78712, USA}

\begin{abstract}
\noindent
Recent progress regarding the noncanonical Hamiltonian formulation of extended magnetohydrodynamics (XMHD), a model with Hall drift and electron inertia, is summarized. The advantages of the Hamiltonian approach are invoked to study some general properties of XMHD turbulence, and to compare them against their ideal MHD counterparts. For instance, the helicity flux transfer rates for XMHD are computed, and Liouville's theorem for this model is also verified. The latter is used, in conjunction with the absolute equilibrium states, to arrive at the spectra for the invariants, and to determine the direction of the cascades, e.g., generalizations of the  well-known  ideal MHD inverse cascade of magnetic helicity. After a similar analysis is conducted for XMHD by inspecting second order structure functions and absolute equilibrium states, a couple of interesting results emerge. When cross helicity is taken to be ignorable, the inverse cascade of injected magnetic helicity also occurs in the Hall MHD range - this is shown to be consistent with previous results in the literature. In contrast, in  the inertial MHD range, viz.\  at scales smaller than the electron skin depth, all spectral quantities are expected to undergo direct cascading. The consequences and relevance of our results in space and astrophysical plasmas are also briefly discussed.
\end{abstract}

\maketitle

\section{Introduction}
In most areas of fusion, space and astrophysical plasmas, fluid models have proven to be highly useful in capturing the relevant physics \citep{Dung58,KT73,GP04,Fre14}. Amongst them, the  simplest and most widely used is ideal magnetohydrodynamics (MHD). Although MHD has proven to be very successful in predicting many phenomena, it is known to be valid only in certain regimes. There exist a wide class of systems, particularly in astrophysics and space science, which are collisionless with non-ideal MHD effects becoming important. For instance, one such notable contribution is the Hall effect that  becomes non-negligible when the characteristic frequencies become comparable to, or greater than, the ion cyclotron frequency $\omega_{ci}$ \citep{bellan08}. Another crucial effect worth highlighting is due to electron inertia, which becomes important when one considers characteristic length scales that are smaller than the electron skin depth $d_e = c/\omega_{pe}$ with $\omega_{pe}$ denoting the electron plasma frequency. 

Thus, it is advantageous to seek fluid models containing  the above two effects. Extended MHD, henceforth referred to as XMHD, is a model that is endowed with both the Hall drift and electron inertia \citep{Lust59}. It can be rigorously derived from two-fluid theory through a series of systematic orderings and expansions, as shown in \citet{Dung58,GP04,KLMWW14}. Although it has long since been known that ideal MHD has both action principle \citep{Newcomb62} and Hamiltonian \citep{MG80} formulations, the XMHD equivalents proved to be quite elusive until recently - the former was presented in \citet{KLMWW14,DAML16} and the latter in \citet{AKY15,LMM15}. At this stage, it is important and instructive to pose two crucial questions. What general benefits do the Hamiltonian and Action Principle (HAP) formulations accord? Secondly, what are the physical systems and phenomena where extended MHD has been successfully employed?

The first question has already been explored extensively, and we refer the reader to the reviews by \citet{serrin,Mor82,ZMR85,HMRW85,S88,ZLF92,M98,Mor05}. Some of the chief advantages, apart from their inherent mathematical elegance and simplification, include:
\begin{itemize}
\item A systematic and rigorous means of constructing equilibria and obtaining sufficient conditions for their stability \citep{HMRW85}. This was recently applied to ideal MHD in a series of works by \citet{AMP13,AMP16}; see also \citet{MTT13}.
\item A clear derivation of reduced models without the loss of the Hamiltonian nature, and thereby avoiding `spurious' dissipation; for e.g., \citet{MH84,HHM87,MLA14,Lin15,KWM15}.
\item The extraction of important invariants such as the magnetic helicity and its generalizations \citep{Cal63,HMRW85,PMor96a,LMM16}. This is done by means of the particle relabelling symmetry \citep{PMor96a} in the action principle approach and via the degeneracy of the noncanonical Poisson bracket in the Hamiltonian formulation \citep{YMD14}. It is also possible to establish and  elucidate topological properties of XMHD by means of the HAP approach,  as recently shown in \citet{DAML16,LMM16}. 
\item A detailed understanding of how magnetic reconnection operates by taking advantage of the underlying Hamiltonian structure, such as the aforementioned invariants \citep{OP93,CGPPS,GCPP01,TMWG,CGTW12,CGWB13,HHM15}.
\item A natural means of arriving at weak turbulence theories, as described in \citet{ZLF92,ZK97,Naz11}.  This methodology was applied to Hall MHD by \citet{SBR03}.  The reader is directed to the analysis by \citet{ALM16} that drew extensively upon the HAP approach (for e.g., to construct nonlinear wave solutions), and thereby arrived at the energy and helicity spectra of XMHD.  We also point out the recent beatification procedure of \citet{thiago} as an elegant alternative, that explicitly relies on the Hamiltonian formulation.
\item The knowledge of the HAP structures has proven to be highly useful numerically for constructing structure preserving  integrators (variational and symplectic) \citep{CR91,QG08,ES13,EFQT,TBS16,KKMS16}. These integrators have (definitively) outperformed other conventional choices, as the latter lack the unique conservation laws and geometric properties of the former.
\end{itemize}
In addition to these (admittedly representative) benefits, we also observe that the HAP approach has been tangentially employed in astrophysical phenomena such as Hall MHD dynamos \citep{PDM03,LinB16,LB16} and jets \citep{LMa15}. Thus, it is quite evident that a thorough understanding of the Hamiltonian and Action Principle (HAP) formalisms for XMHD is quite warranted. However, this brings us to the second point, concerning the \emph{physical} relevance and importance of the model (XMHD) itself. 

Fortunately, there are several instances where XMHD has proven to be a very useful physical model. From the perspective of fundamental plasma phenomena, both turbulence and reconnection results have been radically altered since  the Hall term (and electron inertia) was taken into account. In the case of the latter field in particular, it is not an exaggeration to say that the whole field was revitalized through the inclusion of this one simple term. The reader may consult the excellent texts by \citet{Bisk00,BP07} on this subject. In turbulence, it has been shown that the introduction of Hall drift (and electron inertia) leads to the steepening of spectra \citep{KM04,Gal06,GB07,SS09,ALM16}. Each of these theoretical consequences has  been confirmed through detailed observations of the Earth's magnetosphere \citep{KR95,Butal16}, and the solar wind and corona \citep{SGRK09,PDM12,BC13,SHB13,ACSHB13,Nar16}. Lastly, we also wish to note that certain fusion phenomena, such as sawtooth crashes \citep{Hast97}, have also been explained well by utilizing XMHD. 
 
The outline of our paper is as follows. In Sec.~\ref{SecHam}, we discuss the noncanonical Hamiltonian structure of XMHD, and some of its salient features. We follow this up by using some of these aspects to gain a better understanding XMHD turbulence. In Sec.~\ref{nonisotropxmhd}, we generalize the results of \citet{BG16} on helicity flux transfer rates to include electron inertia, and we also briefly raise the issue of the directional nature of the cascades. This matter is addressed in more detail in Sec.~\ref{dircascades}, where we predict that the inverse cascade of magnetic helicity operates in the Hall MHD regime, but is absent when we consider the inertial MHD range that is valid at sub-electron skin depth scales. Finally, we conclude with a summary of our results and their implications in Sec.~\ref{SecConc}.

 
 \section{The Hamiltonian formulation of extended MHD} \label{SecHam}
 
In the present section, we introduce the equations of the extended MHD model, and discuss its Hamiltonian structure as well as the insights that follow as a natural consequence.  

\subsection{Preliminary model considerations}

Although the correct form of the equations of XMHD has been known since the 1950s \citep{Dung58,Lust59}, many different variants exist in the literature. Of these, it is worth remarking that some of them are incorrect and do not conserve  energy \citep{KM14}.

The XMHD equations comprise of the continuity equation, the momentum equation and the generalized Ohm's law \citep{GP04,BP07}. They are respectively given by 
\begin{equation} \label{ContEq}
\frac{\p \rh}{\p t} + \nabla \cdot \left(\rh {\bf V}\right) = 0,
\end{equation}
\begin{equation} \label{MomDensv1}
\rh \left(\frac{\p {\bf V}}{\p t} + {\bf V}\cdot \nabla {\bf V}\right) = - \nabla p + {\bf J} \times {\bf B} 
\,- \frac{m_e}{e^2} {\bf J}\cdot \nabla \left(\frac{{\bf J}}{n}\right), 
\end{equation}
\begin{equation} \label{Ohmv1}
{\bf E} + {\bf V} \times {\bf B} - \frac{{\bf J} \times {\bf B} - \nabla p_e + \mu \nabla p_i}{en}
= \frac{m_e}{ne^2} \left[\frac{\p {\bf J}}{\p t} + \nabla \cdot \left({\bf V} {\bf J} + {\bf J} {\bf V} - \frac{1}{en} {\bf J} {\bf J} \right) \right].
\end{equation}
Here, note that the one-fluid variables $\rh$, ${\bf V}$ and ${\bf J} = \mu_0^{-1} \nabla \times {\bf B}$ are the total mass density, the centre-of-mass velocity and the current respectively. ${\bf E}$ and ${\bf B}$ denote the electric and magnetic fields, whilst $p_s$ is the pressure of species `$s$' and $p = p_i + p_e$ is the total pressure. The variables $m_e$ and $e$ are the electron mass and charge, whilst $\mu = m_e/m_i$ is the mass ratio. An inspection of (\ref{Ohmv1}) reveals that it is far more complex than the ideal MHD Ohm's law that  follows by setting all terms except the first two (on the LHS) to zero.

The next step is to render the above equations dimensionless. This is done by normalizing everything in terms of Alfv\'enic units, and the reader is directed to \citet{AKY15,LMM16} for further details. We also introduce the dynamical variable
\begin{equation} \label{starB}
{\bf B}^* = {\bf B} + d_e^2 \, \nabla \times \left[\frac{\nabla \times {\bf B}}{\rh} \right],
\end{equation}
which is well known from previous theories that relied upon electron inertia, such as \citet{OP93,CGPPS}. After some algebraic manipulation (\ref{MomDensv1}) and (\ref{Ohmv1}) can be expressed in a simpler manner as follows. 
\begin{equation}  \label{MomDensv2}
\frac{\p {\bf V}}{\p t} + \left(\nabla \times {\bf V}\right) \times {\bf V} = -\nabla \left(h + \frac{V^2}{2}\right) + \frac{\left(\nabla \times{\bf B}\right) \times {\bf B}^*}{\rh} - d_e^2 \nabla \left[\frac{\left(\nabla \times {\bf B}\right)^2}{2\rh^2}\right],
\end{equation}
\begin{equation} \label{Ohmv2}
\frac{\p {\bf B}^*}{\p t} = \nabla \times \left({\bf V} \times {\bf B}^* \right) - d_i \nabla \times \left(\frac{\left(\nabla \times{\bf B}\right) \times {\bf B}^*}{\rh}\right) + d_e^2 \nabla \times \left[\frac{\left(\nabla \times {\bf B}\right) \times \left(\nabla \times {\bf V}\right) }{\rh}\right].
\end{equation}
In obtaining the above two equations, we observe that a barotropic pressure was implicitly assumed; for a non-barotropic treatment, we refer the reader to \citet{KLMWW14,DAML16}. In the above expressions, note that $d_s = c/(\omega_{p s} L)$ is the skin depth of species `$s$' normalized to the characteristic length scale $L$, and $\omega_{p s}$ is the corresponding plasma frequency. All of these values are in terms of the fiducial units that were adopted for the purpose of normalization. 

\subsection{The Hamiltonian structure of extended MHD}

We are now in a position to present the Hamiltonian formulation of extended MHD. A detailed derivation of this structure can be found in \citet{AKY15,LMM15,DAML16} and a recent overview was provided in \citet{LMM16}. 

Firstly, we observe that (\ref{ContEq}), (\ref{MomDensv2}) and (\ref{Ohmv2}) can be used to show that the following energy functional is conserved. 
\begin{equation} \label{HamExtMHD}
H = \int_D d^3x\,\left[\frac{\rh V^2}{2} + \rh U(\rh) + \frac{B^2}{2} + d_e^2 \frac{\left(\nabla \times {\bf B}\right)^2}{2\rh} \right],
\end{equation}
where $D \subset \mathbb{R}^3$ \citep{Mor82}. It is worth remarking that there is no $d_i$ dependence, but there is a $d_e$-dependent term, which stems from the electron fluid velocity. Upon setting $d_e \rightarrow 0$, we will obtain the famous ideal MHD energy \citep{Newcomb62}.

From basic classical mechanics, we know that a system can be rendered Hamiltonian if one has a conserved ``energy'' and a suitable Poisson bracket. The Poisson bracket must satisfy the properties of (i) bilinearity, (ii) antisymmetry, (iii) the Leibniz product rule, and (iv) the Jacobi identity \citep{SM16}. Even though our Poisson bracket is infinite-dimensional (and degenerate), it must still satisfy these properties. 

It is instructive to first begin with Hall MHD (HMHD), which is the best known of all the extended magnetofluid models. As noted earlier, it has proven to be highly useful in explaining phenomena such as magnetic reconnection, dynamos,  and turbulence. The correct noncanonical Poisson bracket was provided by \citet{YH13}, and is given by
\begin{eqnarray} \label{HallMHDNCPB}
\{F,G\}^{HMHD} &=& - \int_D d^3x\,\Bigg\{\left[F_\rh \nabla \cdot G_{\bf V} + F_{\bf V} \cdot \nabla G_\rh \right] - \frac{\left(\nabla \times {\bf V}\right)}{\rh} \cdot \left(F_{\bf V} \times G_{\bf V}\right) \nonumber \\
&& \hspace{0.7 in}  - \frac{{\bf B}}{\rh} \cdot \left(F_{\bf V} \times \left(\nabla \times G_{\bf B}\right) \right)  + \frac{{\bf B}}{\rh} \cdot \left(G_{\bf V} \times \left(\nabla \times F_{\bf B}\right)\right) \Bigg\} \nonumber \\
&&-\, d_i \int_D d^3x\,\frac{{\bf B}}{\rh} \cdot \left[\left(\nabla \times F_{\bf B}\right) \times \left(\nabla \times G_{\bf B}\right)\right], \nonumber
\end{eqnarray}
where $F_{\phi}:= \delta F/\delta \phi$ is the functional derivative \citep{Mor82} with respect to $\phi$. It is more transparent to write the above equation as 
\begin{equation} \label{HMHDOldVar}
\{F,G\}^{HMHD} = \{F,G\}^{MHD} + \{F,G\}^{Hall},
\end{equation}
where the first two lines of (\ref{HallMHDNCPB}) constitute the classic MHD bracket $\{F,G\}^{MHD}$ that was first derived by \citet{MG80}. The last line of (\ref{HallMHDNCPB}), with the factor of $d_i$ in front, gives rise to the Hall contributions in the Ohm's law. 

Next, let us turn our attention to (\ref{Ohmv1}) once more. The Hall MHD Ohm's law follows by setting everything on the RHS alone to zero. Instead, suppose we consider a case where the third term, with a factor of $en$ in the denominator, is set to zero. In our choice of normalized units, this amounts to setting $d_i \rightarrow 0$ but \emph{not} $d_e \rightarrow 0$ as well. This may appear counterintuitive, but we observe that $d_i$ and $d_e$ must be perceived as independent variables. The resultant model has sometimes been referred to as inertial MHD \citep{KM14,LMT14}, because it encompasses electron inertia but not the Hall term. 

Although inertial MHD (IMHD) may appear somewhat \emph{ad hoc} at this stage, it can be derived through a rigorous ordering procedure as discussed in \citet{KM14,LMM16}. Moreover, it is particularly useful in deriving reduced models for reconnection \citep{OP93,FP04,FP07}. It has also proven to be useful in studying dynamo action \citep{Kle94}, as the inertial MHD Ohm's law is linear in ${\bf B}$. In \citet{LMM15,LMM16}, it was shown that a remarkable equivalence between the Hall and inertial MHD Poisson brackets exists. This equivalence can be expressed as
\begin{equation} \label{HallInertEqv}
\{F,G\}^{IMHD} \equiv \{F,G\}^{HMHD} \left[\mp 2d_e;\,\boldsymbol{\calb}_{\pm}^{(I)}\right],
\end{equation}
where the LHS is to be understood as follows. Replace ${\bf B}$ everywhere in the Hall MHD bracket (\ref{HallMHDNCPB}) with $\boldsymbol{\calb}_{\pm}^{(I)}:= {\bf B}^* \pm d_e \nabla \times {\bf V}$ and $d_i$ with $\mp 2d_e$, where ${\bf B}^*$ was defined in (\ref{starB}). Owing to the presence of the `$\pm$', it is clear that there are \emph{two} such transformations which lead to the equivalence. In Sec.~\ref{XMHDTop}, we shall comment on the nature of these transformations further. 

Finally, let us consider extended MHD (XMHD) in its entirety, i.e.,  where no terms are dropped from the Ohm's law (\ref{Ohmv1}). The noncanonical Poisson bracket for this model was derived by \citet{AKY15}, and \citet{LMM15} showed that another beautiful equivalence between the Hall and extended MHD brackets existed. In mathematical terms, it amounts to
\begin{equation} \label{ExHallMHDEqv}
\{F,G\}^{XMHD} \equiv \{F,G\}^{HMHD}\left[d_i-2\kappa_{\pm};\,\boldsymbol{\calb}_\pm \right], 
\end{equation}
where the RHS indicates that the substitutions
\begin{equation} \label{Bhyb}
{\bf B} \rightarrow \boldsymbol{\calb}_\pm:= {\bf B}^* + \kappa_{\pm} \nabla \times {\bf V},
\end{equation}
and $d_i \rightarrow d_i-2\kappa_\pm$ in (\ref{HallMHDNCPB}) lead to the XMHD bracket. Again, there are two such transformations since $\kappa_\pm$ follow from determining the two roots of the quadratic equation
\begin{equation}
    \kappa^2 - d_i \kappa - d_e^2 = 0.
\end{equation}
Here, observe that setting $d_i = 0$ leads us to the inertial-Hall MHD equivalence discussed above, and also transforms (\ref{Bhyb}) to $\boldsymbol{\calb}_{\pm}^{(I)}$.

Before proceeding further, a comment on why these connections between the different models are remarkable is in order. In Hall MHD, there is \emph{no} electron inertia but there is a \emph{finite} Hall drift. In inertial MHD, the situation is exactly reversed, i.e. there is no Hall drift but there is electron inertia. Thus, it is not at all intuitively obvious that the two models could share a common Hamiltonian structure, since their effects are mutually exclusive. Yet, the above relations show that there does exist a deep, and non-trivial, equivalence between the two models. This equivalence is also shared by extended MHD, which has \emph{both} Hall drift and electron inertia. Here, it must be understood that the ``equivalence'' referred to thus far between Hall MHD and inertial MHD is only concerned with their respective Poisson brackets. The corresponding Hamiltonians for these two models are not identical, as they differ by a single term, i.e. the last one in \eqref{HamExtMHD}.

Subsequently, we shall explore the different predictions regarding the behavior of turbulent cascades in Sections \ref{HMHDCas} and \ref{IMHDCas}.

\subsection{On the topological properties of extended MHD}
 \label{XMHDTop}

We have seen earlier that the variable (\ref{Bhyb}) lies at the heart of the equivalence between the different models. To understand why, it is instructive to take a step backwards and consider ideal MHD. In any introductory textbook, the frozen-flux property of ideal MHD and the conservation of magnetic helicity $\int_D d^3x\, {\bf A} \cdot {\bf B}$ are presented. Thus, it is natural to ask if one can seek generalizations of these properties to XMHD, since both of these features are present in two-fluid theory \citep{SK82,SI97} and in Hall MHD \citep{Turn86}.

In ideal MHD, the frozen-flux constraint can be expressed as
\begin{equation}
{\bf B} \cdot d{\bf S} = {\bf B}^0 \cdot d{\bf S}^0,
\end{equation}
where $d{\bf S}$ is the area element, and the superscript `$0$' denotes the values at $t=0$ \citep{Newcomb62}. It is also possible to view the above expression as the statement that the magnetic flux (in ideal MHD) is a Lie-dragged 2-form; for more details, the reader may consult \citet{TY93}.

In XMHD, there are \emph{two} such generalized frozen-flux constraints, given by 
\begin{equation}
\boldsymbol{\calb}_\pm \cdot d{\bf S}_\pm = \boldsymbol{\calb}_\pm^0 \cdot d{\bf S}^0_\pm,
\end{equation}
where $\boldsymbol{\calb}_\pm$ was defined in (\ref{Bhyb}) and $d{\bf S}_\pm$ denotes the corresponding area element. This elegant property was first recognized in \citet{LMM15}, later proven in \citet{LMM16} and utilized further in \citet{DAML16}.

Next, let us consider the helicity. In ideal MHD, the magnetic helicity is conserved, but it is no ordinary invariant. Instead, it is both a Casimir invariant and a topological invariant. Casimir invariants are special invariants that follow from the degeneracy of the (noncanonical) Poisson bracket, and they are found via $\{F,C\} = 0 \,\,\forall\, F$, with $C$ denoting the Casimir invariant. They play an important role in regulating the phase space dynamics, as discussed in \citet{M98}, and have played an important role in reconnection over the years \citep{OP93,CGPPS,GCPP01}. Magnetic helicity is also a topological invariant since it is closely connected with the linking and twisting of field lines - more precisely, it shares close connections with the Gauss linking number as discussed in \citet{Mor61,M69,BF84}.

Thus, one can obtain the generalized counterparts of the magnetic helicity in extended MHD by seeking out the Casimir invariants that resemble it. There are two such invariants 
\begin{equation}\label{helicities}
	\mathcal{K}_\pm = \int_D d^3 x\,\boldsymbol{\mathcal{A}}_\pm \cdot
	\boldsymbol{\mathcal{B}}_\pm,
\end{equation}
where $\boldsymbol{\mathcal{B}}_\pm = \nabla \times \boldsymbol{\mathcal{A}}_\pm$, and the LHS is given by (\ref{Bhyb}). It is clear that these generalized helicities have the same form of the magnetic and fluid helicities (for MHD and HD respectively), and hence one may expect them to share similar topological properties. This conjecture was confirmed in \citet{LMM16}, where the authors also established some unusual connections with Chern-Simons theory, a ubiquitous (topological) quantum field theory that appears in high-energy and condensed matter physics. 

From the preceding discussion, it is clear that the variable $\boldsymbol{\mathcal{B}}_\pm$ that facilitates the equivalence between the different extended magnetofluid models is \emph{not} arbitrary. It has close connections with the generalized frozen-fluxes, helicities and Lie-dragged 2-forms all of which have clear mathematical and physical significance. Lastly, it is also possible to manipulate (\ref{MomDensv2}) and (\ref{Ohmv2}) directly to arrive at
\begin{equation} \label{GenIndRel}
	\partial_t \boldsymbol{\mathcal{A}}_\pm =
	\mathbf{V}_\pm\times\boldsymbol{\mathcal{B}}_\pm + \nabla \psi_\pm
	\quad\text{and}\quad
	\partial_t \boldsymbol{\mathcal{B}}_\pm = 
	\nabla\times(\mathbf{V}_\pm\times\boldsymbol{\mathcal{B}}_\pm ),
\end{equation}
where $	\mathbf{V}_\pm:=\boldsymbol{V}-\kappa_\mp \nabla\times\boldsymbol{B}$ and
\begin{equation} \label{Psidefn}
	\psi_\pm:=\kappa_\mp h_e-\Big(\kappa_\pm+\frac{d_e^2}{d_i}\Big) h_i - \phi + \kappa_\mp
	d_e^2\, \frac{J^2}{2\rho}-d_e^2\frac{\mathbf{J}\cdot\mathbf{V}}{\rho},
\end{equation}
as shown in \citet{LMM16}. Upon inspection, it is clear that the second set of equations in (\ref{GenIndRel}) exactly resemble the induction equation in ideal MHD, thereby emphasizing the role of $\boldsymbol{\mathcal{B}}_\pm$ as the generalization of the magnetic field. It is, however, more common to refer to it as the generalized (or canonical) vorticity. 

Thus, to summarize our discussion up to this point, we have seen that the Hamiltonian formulation of XMHD has led us to two important conclusions.
\begin{itemize}
\item There exists a high degree of mathematical similarity between the different models, even though they have contrasting (and sometimes exclusive) physical effects. This mathematical equivalence between the models is rendered very clear when written in  Hamiltonian form. Hence, the latter approach serves as a means of unifying the different extended MHD models. 
\item The similarities between extended MHD and ideal MHD can be understood further by means of the HAP formulations, which lead us to the generalizations of the helicity, flux,  and induction equation. 
\end{itemize}
Bearing these advantages in mind, we shall now proceed to study some pertinent features of XMHD turbulence in the subsequent sections. 
 
 
\section{Flux transfer rates}
\label{nonisotropxmhd}

In the recent work by \citet{BG16},  expressions for the dissipation rates for Hall MHD were computed. Their analysis assumed that the Hall MHD turbulence was homogeneous, but did not rely on the further assumption of isotropy. As noted above, an important limitation of Hall MHD is that it becomes invalid when electron inertia effects start to dominate, i.e.\  when one considers length scales comparable to the electron skin depth. In such an instance, it makes sense to use extended MHD instead, on account of the fact that it is endowed with electron inertia effects. 

In recent times, there has also been a great deal of interest focused on  the solar wind at sub-electron scales, mostly because of the fact that observations have now become possible in this regime \citep{SGRK09,Alex09,Set11,BC13,SHB13}. Hence, in the present section, we shall generalize the results of \citet{BG16} by including electron inertia.

\subsection{Mean helicity flux rates}

\begin{figure}[h]
	\begin{tikzpicture}
	\foreach \x in {-6,-4,...,6}	
		\draw[red,<-,very thick] (\x,0) arc (0:180:1);
	\draw[->,very thick]  (-8.5,-.1) -- (7,-.1) node[very near start,anchor=north] {Driving Range} node[midway,anchor=north] {Inertial Range} node[very near end,anchor=north] {Dissipative Range} node[anchor=south] {$k$};
	\end{tikzpicture}
	\caption{\small Schematics of the standard Richardson-Kolmogorov direct cascade. Energy injected at  low $k$,  e.g.\  via large scale stirring, cascades through the inertial range and dissipates at small scales (large $k$). Upon reversal of the arrows along with the driving and dissipative ranges, the mechanism of the inverse cascade is obtained.}\label{range}
\end{figure}
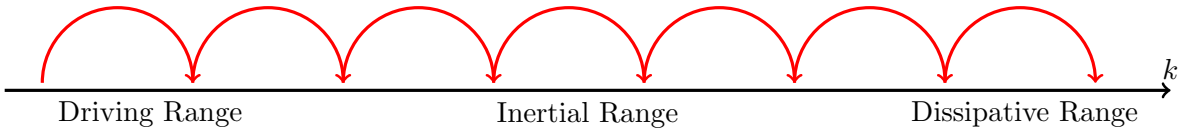

In 3D fluid turbulence, it has been known since the famous works by \citet{Kol41}, and subsequent numerical and experimental tests \citep{frisch95,SA97,biskamp03}, that the energy input at large scales flows to  small dissipative scales. This phenomenon is often referred to as a direct Kolmogorov-Richardson cascade \citep{frisch95} - a pictorial description of this phenomenon has been provided in Fig.~\ref{range}. 

In MHD, the direct cascade of energy and the inverse cascade of magnetic helicity \citep{frisch75} have been widely explored, and are thus well established \citep{biskamp03}. In the inertial range, it must be borne in mind that the dissipation does not play a role. Hence, it is expected that, in the stationary regime, the same flux (of the energy or helicity, for example) flows through each wave number $k$. This principle was recently employed to conduct a complementary study of XMHD turbulence in \citet{ALM16}.

In our analysis, we are interested in the flux rate of the generalized helicities (\ref{helicities}) within the framework of XMHD that are injected at some length scale. By following the steps outlined in \citet{BG16}, we first introduce the symmetric two-point correlation function
\begin{equation}
R_{\mathcal{K}_{\pm}} = R_{\mathcal{K}_{\pm}}' = \Bigg\langle \frac{\boldsymbol{\mathcal{A}}^\prime_\pm \cdot
	\boldsymbol{\mathcal{B}}_\pm + \boldsymbol{\mathcal{A}}_\pm \cdot
	\boldsymbol{\mathcal{B}}^\prime_\pm}{2} \Bigg\rangle,
\end{equation}
where  primed quantities are functions of $\mathbf{x}^\prime = \mathbf{x}+\mathbf{r}$,    unprimed quantities  depend on $\mathbf{x}$, and the  brackets $\langle\ \rangle$ are a shorthand notation for ensemble averaging. When the turbulence is homogeneous this can be equivalent to the spatial average. Upon manipulation  we find  
\begin{eqnarray} 
\label{ABCorF}
	&&\partial_t\big\langle\boldsymbol{\mathcal{A}}^\prime_\pm \cdot
	\boldsymbol{\mathcal{B}}_\pm + \boldsymbol{\mathcal{A}}_\pm \cdot
	\boldsymbol{\mathcal{B}}^\prime_\pm \big\rangle  =  \big\langle
	\nabla\cdot\lbrack(\mathbf{V}_\pm\times\boldsymbol{\mathcal{B}}_\pm)
	\times\boldsymbol{\mathcal{A}}^\prime_\pm \rbrack+  
	\nabla^\prime\cdot\lbrack(\mathbf{V}_\pm^\prime\times\boldsymbol{\mathcal{B}}_
	\pm^\prime)\times\boldsymbol{\mathcal{A}}_\pm \rbrack 
	\nonumber\\
	&& \hspace{ 2cm}
	 + \  \mathbf{V}_\pm^\prime\times\boldsymbol{\mathcal{B}}_
	\pm^\prime\cdot\boldsymbol{\mathcal{B}}_\pm +
	\mathbf{V}_\pm\times\boldsymbol{\mathcal{B}}_
	\pm\cdot\boldsymbol{\mathcal{B}}_\pm^\prime
	+\nabla\psi_\pm\cdot\boldsymbol{\mathcal{B}}_\pm^\prime
	 + \nabla^\prime\psi_\pm^\prime\cdot\boldsymbol{\mathcal{B}}_\pm \big\rangle\,.
\end{eqnarray}
At this stage, we shall digress a little to explain how the principle of statistical homogeneity can be gainfully employed. Our discussion mirrors the one presented in \citet{george13}. We introduce the change-of-variables $\boldsymbol{\xi} = \mathbf{x}^\prime+\mathbf{x}$ and $\mathbf{r} = \mathbf{x}^\prime - \mathbf{x}$, which implies that $\partial/\partial\mathbf{x} = \partial/\partial\boldsymbol{\xi} -\partial/\partial\boldsymbol{r} $ and $\partial/\partial\mathbf{x}^\prime =
\partial/\partial\boldsymbol{\xi}+\partial/\partial\mathbf{r}$. For a given vector field ${\bf u}$, this implies that
\begin{eqnarray}
	\Big\langle u_j(\mathbf{x}^\prime,t)\frac{\partial u_i(\mathbf{x},t)}{\partial
	x_i} \Big\rangle &=& \frac{\partial}{\partial x_i}\langle
	u_j(\mathbf{x}^\prime,t) \,u_i(\mathbf{x},t) \rangle =
	 -\frac{\partial}{\partial r_i}\langle u_j(\mathbf{x}^\prime,t)
	\,u_i(\mathbf{x},t) \rangle \nonumber\\
	&=& -\frac{\partial}{\partial x_i^\prime}\langle
	u_j(\mathbf{x}^\prime,t) \,u_i(\mathbf{x},t) \rangle =
	- \Big\langle \frac{\partial u_j(\mathbf{x}^\prime,t)}{\partial
	x_i^\prime} u_i(\mathbf{x},t) \Big\rangle\,.
\end{eqnarray}
Upon using the above identity in (\ref{ABCorF}), we obtain
\begin{eqnarray}
	&&\big \langle\nabla\cdot\lbrack(\mathbf{V}_\pm\times\boldsymbol{\mathcal{B}}_\pm)
	\times\boldsymbol{\mathcal{A}}^\prime_\pm \rbrack + 
	\nabla^\prime\cdot\lbrack(\mathbf{V}_\pm^\prime\times\boldsymbol{\mathcal{B}}_
	\pm^\prime)\times\boldsymbol{\mathcal{A}}_\pm \rbrack\big \rangle \nonumber\\
	&& \hspace{1.5 cm} = -\,\big\langle\nabla^\prime\cdot\lbrack(\mathbf{V}_\pm
	\times\boldsymbol{\mathcal{B}}_\pm)
	\times\boldsymbol{\mathcal{A}}^\prime_\pm \rbrack + 
	\nabla\cdot\lbrack(\mathbf{V}_\pm^\prime\times\boldsymbol{\mathcal{B}}_
	\pm^\prime)\times\boldsymbol{\mathcal{A}}_\pm \rbrack \big\rangle \nonumber\\
	&&  \hspace{1.5 cm}  = \big \langle\mathbf{V}_\pm\times\boldsymbol{\mathcal{B}}_\pm
	\cdot\boldsymbol{\mathcal{B}}^\prime_\pm  + 
	\mathbf{V}_\pm^\prime\times\boldsymbol{\mathcal{B}}_
	\pm^\prime\cdot\boldsymbol{\mathcal{B}}_\pm  \rangle = -\langle\delta
	(\mathbf{V_\pm\times\boldsymbol{\mathcal{B}}_\pm})\cdot\delta
	\boldsymbol{\mathcal{B}}_\pm \big\rangle\,,
\end{eqnarray}
where $\delta f := f^\prime - f$. Likewise, it is possible to show that
\begin{equation}
\big\langle\nabla^\prime\psi^\prime_\pm\cdot\boldsymbol{\mathcal{B}}_\pm
+\nabla\psi_\pm\cdot\boldsymbol{\mathcal{B}}_\pm^\prime \big\rangle =
-\big\langle\psi^\prime_\pm\nabla\cdot\boldsymbol{\mathcal{B}}_\pm
+\psi_\pm\nabla^\prime\cdot\boldsymbol{\mathcal{B}}_\pm^\prime\big\rangle = 0.
\end{equation}
Thus, upon combining everything together, we get
\begin{equation}\label{heldissipation}
	\frac{\partial}{\partial
	t}\Big\lbrack\frac{1}{2}\,\big\langle\boldsymbol{\mathcal{A}}^\prime_\pm \cdot
	\boldsymbol{\mathcal{B}}_\pm + \boldsymbol{\mathcal{A}}_\pm \cdot
	 \boldsymbol{\mathcal{B}}^\prime_\pm \big\rangle\Big\rbrack = -\big\langle\delta
	(\mathbf{V_\pm\times\boldsymbol{\mathcal{B}}_\pm})\cdot\delta
	\boldsymbol{\mathcal{B}}_\pm\big\rangle + D\,,
\end{equation}
where we have introduced the phenomenological damping $D$ that occurs at the sink scale, following the approach of \citet{BG16}. In the limit of infinite kinetic and magnetic Reynolds numbers, under the assumption of stationarity, the LHS of the above expression vanishes due to the ruggedness of the helicity invariants \citep{MG82}. Hence, the large scale dissipation equals the mean generalized helicity flux rate 
\begin{equation} \label{HelDissRate}
\eta_{\pm} = \big\langle\delta
 (\mathbf{V_\pm\times\boldsymbol{\mathcal{B}}_\pm})\cdot\delta
 \boldsymbol{\mathcal{B}}_\pm\big\rangle\,, 
\end{equation}
which closely resembles  the expression of \citet{BG16}. However, it must be noted  that our expression is more general as it duly encompasses electron inertial contributions as well via the definition of $\boldsymbol{\mathcal{B}}_\pm$. In the Hall MHD limit with $d_e \rightarrow 0$, we have verified that our result is in exact agreement with  the expression of \citet{BG16}.

Although (\ref{HelDissRate}) is quite compact, a great deal of information can be extracted from it. For instance, it follows that the dissipation rates vanish when the Beltrami condition $\boldsymbol{\mathcal{B}}_\pm \parallel \mathbf{V}_\pm$ is attained. These (multi) Beltrami states are non-trivial, as they are also  equilibria of XMHD \citep{ALM16}. This is easy to verify by inspecting the second set of equations in (\ref{GenIndRel}), and substituting the above condition. Thus, this result serves as a consistency check indicating that the dissipation vanishes when the system has settled into this equilibrium (in the limit of infinite Reynolds numbers). 

In \citet{BG16} a phenomenological argument for the direction of the cascades was presented. First, let us recall that the generalized helicities become the magnetic and ion canonical helicities in Hall MHD \citep{Turn86}. The first is essentially a copy of the MHD magnetic helicity, while the other is a superposition of MHD cross helicity and fluid helicity after some rearrangement. In the former, it is argued that the inverse cascade is expected just as in ideal MHD. In contrast, the direction of the cascade for the ion canonical helicity (of Hall MHD) can go either way, as it is dependent on the energy budget of the system. It is assumed to exhibit an inverse cascade if the magnetic energy is dominant over the kinetic (and thermal) energy. 

Therefore, we see that there is an ambiguity regarding the directionality of the cascade for one of the helicities. The problem becomes far more acute when we include electron inertia effects via XMHD. In that case, the magnetic helicity is not conserved as there is also a (smaller) fluid helicity contribution. If we apply the above line of reasoning, we would expect to witness the direct and inverse cascades of both helicities in XMHD. This is because of the fact that the two helicities are not fundamentally different, other than the fact that they are associated with different species \citep{LMM15,LMM16}. Thus, this raises an interesting question: how is it possible to get the Hall MHD limit from XMHD? In other words,  why is the direct cascade of one helicity, that corresponds to the magnetic helicity in the HMHD limit, lost? One possible resolution of this paradox is by suggesting that the existence of direct or inverse cascades depends on the length scale we are considering. This question is addressed in more detail in Sec.~\ref{dircascades} that follows. 


\section{Direction of Cascades}
\label{dircascades}

\subsection{Liouville's theorem for XMHD}

The direction of a cascade can be determined by inspecting the general equilibrium states that the turbulence would tend  to relax to, if not for the continual input of energy \citep{biskamp03}. Although turbulence as a phenomenon is far from equilibrium,  absolute equilibria have been used to predict the direction of the spectral flux \cite{biskamp03}. Such equilibria can be obtained from the ideal invariants described in Sec.~\ref{XMHDTop}.  The approach delineated in the present section  is a generalization of the pioneering studies in hydrodynamic \citep{lee52,kraichnan67,kraichnan73} and MHD \citep{lee52,frisch75} turbulence. It is important to bear in mind the fact that the resultant equilibrium spectra are merely tools for predicting direction of cascades, and are far removed from the expected spectra of Kolmogorov type \citep{kraichnan73,frisch95,biskamp03}. For the treatment of the latter issue in XMHD, the reader may refer to \citet{ALM16} instead.

However, before applying equilibrium statistical mechanics to the Fourier modes of XMHD, it is necessary to show that their governing equations  satisfy Liouville's theorem, as was first done in hydrodynamics by \citet{burgers33}. It is then possible to apply the conventional assumption of equal \emph{a priori} probabilities in phase space $(z_1, z_2,\dots,z_n)$ \citep{LL80}, which in turn enables one to express an equilibrium  phase space probability density $\mathscr{P} = \mathscr{P}(z_1, z_2,\dots,z_n)$ as a function of constants of motion; for XMHD, they are further discussed in Sec. \ref{abseqstates}. Liouville's theorem was reproven and used for 2D fluids by \citet{kraichnan80}, quasi-geostrophy by \citet{salmon76},  incompressible MHD by \citet{lee52}, and more recently similar statistical approaches have been employed in  plasma models \citep{Krom02}, such as two-fluid theory \citep{ZWZ14} and gyrokinetics \citep{ZhHa10}.

For an $N$-dimensional dynamical system $\dot z_i=V_i(z)$, for some vector field $V$,  with $i=1,2,\dots N$, Liouville's theorem (e.g.\ \cite{khinchin}) states that any phase space volume is preserved provided $ \sum_i \partial  \dot z_i/\partial z_i =\sum_i \partial V_i/\partial z_i=0$, which is true for any canonical Hamiltonian system.
However,  incompressible XMHD is  a noncanonical Hamiltonian system, which  can be  shown \footnote{We shall defer a detailed exposition of  subtleties regarding the noncanonical Hamiltonian origin of the present measure and comparison to an actual canonical measure  to a future publication. On a related note, we also wish to correct an erroneous statement in \citet{kraichnan80} - it was stated therein that 2D fluid flow is not Hamiltonian, but the authors were unaware that it actually is a Hamiltonian dynamical system, albeit in terms of noncanonical variables \citep{M98}.} through the use of Dirac brackets \citep{Dir58}.   Because  Liouville's theorem is variable dependent and the natural (Eulerian) variables are noncanonical,  one must check its validity directly.  The idea of Burgers and Lee was to do this in terms of   Fourier amplitudes, which play the role of  the particle degrees of freedom of statistical mechanics.  Thus, for XMHD  we  write  the  system \eqref{MomDensv2} and \eqref{Ohmv2}, after assuming incompressibility,  in terms of the coefficients of a Fourier series; i.e., the  velocity and magnetic fields are expanded as  $\boldsymbol{F}(\boldsymbol{x})=\sum_{\boldsymbol{k}}
\boldsymbol{f}_k(t) \, e^{i\boldsymbol{k}\cdot\boldsymbol{x}}$. Then the  equations of motion for the Fourier  amplitudes are given by  
\begin{equation}
		\dot {\bs{v}}_{\bs{k}} = i\, \Big(\bs{I} - \frac{\bs{k}\,\bs{k}}{k^2}\Big)\cdot \sum_{\bs{k}^\prime}\Bigg( \bs{v}_{\bs{k}-\bs{k}^\prime}\times \lbrack \bs{k}^\prime \times \bs{v}_{\bs{k}^\prime} \rbrack - \frac{\bs{b}^*_{\bs{k}-\bs{k}^\prime}\times \lbrack \bs{k}^\prime \times \bs{b}^*_{\bs{k}^\prime} \rbrack}{1+{k^\prime}^2 d_e^2} \Bigg),
	\end{equation}	
where $k^2 = \bs{k}\cdot \bs{k}$ and the gradient terms were eliminated via $\nabla\cdot\bs{V} = 0$, and
	\begin{eqnarray}
	{\dot {\bs{b}}_{\bs{k}}}^* &=&  \sum_{\bs{k}^\prime}\Bigg(i  \,\bs{k}\times \lbrack \bs{v}_{\bs{k}^\prime} \times \bs{b}^*_{\bs{k}-\bs{k}^\prime} \rbrack - \frac{d_i\, \bs{k}\times\big\lbrack\bs{b}^*_{\bs{k}-\bs{k}^\prime}\times \lbrack \bs{k}^\prime \times \bs{b}^*_{\bs{k}^\prime} \rbrack\big\rbrack}{1+{k^\prime}^2 d_e^2} 
	\\ \nonumber 
 &&\hspace{ 1cm}  +\  
	\frac{i\, d_e^2}{1+{k^\prime}^2 d_e^2} \big\lbrack \bs{k}\times\bs{k}^\prime\; \bs{v}_{\bs{k}-\bs{k}^\prime}\cdot\bs{k}^\prime \times \bs{b}^*_{\bs{k}^\prime} + \bs{k}\times \bs{v}_{\bs{k}-\bs{k}^\prime}\; \bs{b}^*_{\bs{k}^\prime} \cdot \bs{k}\times \bs{k}^\prime  \big\rbrack  \Bigg).
	\end{eqnarray}
Notice that $\bs{k}\cdot \bs{v}_{\bs{k}} = 0 = \bs{k}\cdot \bs{b}^*_{\bs{k}} $. Technically, our phase space consists of real and complex parts of the vectors $\bs{v}_{\bs{k}} = \overline{\bs{v}}_{-\bs{k}}$ and $\bs{b}^*_{\bs{k}} 
= \overline{\bs{b}^*}_{-\bs{k}}$, where the overbar denotes complex conjugation. However, it is more straightforward to work with their linear combinations $(\bs{v}_{\bs{k}},\overline{\bs{v}_{\bs{k}}},\bs{b}^*_{\bs{k}},\overline{\bs{b}^*_{\bs{k}}})$, and the same results are obtained. After some algebra one arrives at
	\begin{equation}
		\sum_{l,\bs{k}} \frac{\p \dot v_{l\bs{k}}}{\p v_{l \bs{k}}} = -2 i\sum_{\bs{k}}\bs{k}\cdot  \bs{v}_{\bs{0}} = 0,
		\label{vLio}
	\end{equation}
where $l$ indexes the components of $\bs{v}_{\bs{k}}$ and $\bs{v}_{\bs{0}}$ denotes the $k=0$ Fourier component. Even if this component is present, the sum is still zero since it is odd in $\bs{k}$. Similarly, we get
	\begin{equation}
	\sum_{l ,\bs{k}} \frac{\p \dot {b^*}_{l  \bs{k}}}{\p {b^*}_{l \bs{k}}} = -2 \sum_{\bs{k}}\Big(  i\,\bs{k}\cdot  \bs{v}_{\bs{0}} + \frac{d_i \,\bs{b}^*_{\bs{0}} - i \, d_e^2\, \bs{k}\times\bs{v}_{\bs{0}}}{1+k^2 d_e^2}\cdot \bs{k}\times\bs{k}\Big) = 0.
	\label{bLio}
	\end{equation}
Thus, clearly the sum of \eqref{vLio} and \eqref{bLio} vanishes so we have shown that Liouville's theorem holds true in XMHD. Taking the appropriate limits, it is easy to verify that it also holds true for Hall MHD, electron MHD,  and inertial MHD as well. It must be recognized that several past studies of Hall and electron MHD turbulence implicitly relied upon the assumption that Liouville's theorem was valid, without having verified it explicitly. To the best of our knowledge, we have verified it for the first time for XMHD and its simpler variants. 

\subsection{Absolute Equilibrium States}\label{abseqstates}
	
In principle, one can proceed to calculate a partition function for absolute equilibria by using the Hamiltonian and  the two invariants of XMHD, given by ~\eqref{helicities}. However,  because we wish to compare our results with those in the literature by taking the MHD limit, viz. $d_i\rightarrow 0$ and $d_e\rightarrow 0$, and because the generalized  helicities of (\ref{helicities}) become degenerate in this limit, reducing to the magnetic helicity in both instances with a loss of the cross helicity,   it is convenient to use linear combinations of the  helicities (\ref{helicities}). Thus we consider the following two Casimirs: 
	\begin{eqnarray}
		H_M &:=&\frac{1}{2}\frac{\kappa_+ \mathcal{K}_- - \kappa_- \mathcal{K}_+}{\kappa_+ -
		\kappa_-}= \frac{1}{2}\int d^3 x \,(\boldsymbol{A}^*\cdot\boldsymbol{B}^* + d_e^2\,\boldsymbol{V}\cdot\nabla\times\boldsymbol{V}),
		\label{HM}
\\
H_C&:=& \frac{1}{2}\frac{\mathcal{K}_+ - \mathcal{K}_-}{\kappa_+ - \kappa_-}
		=\int d^3 x \,(\boldsymbol{V}\cdot\boldsymbol{B}^* + \frac{d_i}{2}\,\boldsymbol{V}\cdot\nabla\times\boldsymbol{V}),
		\label{HC}
	\end{eqnarray} 
where \eqref{HM}  was also presented in \citet{AKY15,ALM16}. The helicities  \eqref{HM} and \eqref{HC} are natural generalizations of the cross and magnetic helicities of ideal MHD  where the second terms in each of these relations can be seen as ``corrections'' that vanish in the MHD limit.  Here, we have used incompressibility - also a common assumption in most Hall MHD studies \citep{KM04,Gal06} - ensuring that the two dynamical fields are solenoidal in nature.  In  Fourier series representation the  three invariants become 
\begin{eqnarray}
H &=& \frac{1}{2}\sum_{l , \boldsymbol{k}}\Big(
		v_{l \boldsymbol{k}}\overline{v}_{l \boldsymbol{k}} +
		\frac{b^*_{l\boldsymbol{k}} \overline{b^*}_{l \boldsymbol{k}}}{1+k^2 d_e^2}
		\Big),
\\
H_M &=& \frac{i}{2}\sum_{l,m,n,\boldsymbol{k}}\epsilon_{lmn}k_l\Big(
		d_e^2 \,v_{m \boldsymbol{k}}\overline{v}_{n \boldsymbol{k}} +
		\frac{b^*_{m\boldsymbol{k}} \overline{b^*}_{n\boldsymbol{k})}}{k^2}
		\Big),
\\
H_C &=& \frac{1}{2}\sum_{l,\boldsymbol{k}}\Big(
		v_{l\boldsymbol{k}} \overline{b^*}_{l\boldsymbol{k}} +
		b^*_{l\boldsymbol{k}}\overline{v}_{l\boldsymbol{k}}
		+i \sum_{m,n} \,d_i\,\epsilon_{lmn}k_l v_{m\boldsymbol{k}}\overline{v}_{n\boldsymbol{k}}
		\Big).
	\end{eqnarray}
Notice that the energy as well as the helicities are quadratic in ${\bf v}$ and ${\bf b}^*$. At this stage, a few important remarks are in order. Firstly, the statistical mechanics of a large \emph{finite number} of $k$-modes is considered. But, at a later stage, the implicit continuum limit will be taken. Second, we wish to reiterate the central arguments presented in \citet{kraichnan80}. Equilibrium statistical mechanics of classical fields usually results in the ultraviolet (UV) catastrophe. As a result, the analysis that follows is valid only if the system is truncated. This is done by choosing an appropriate cutoff parameter such that $k < k_{max}$. In any realistic system, the dissipation scale yields a natural cutoff, thereby preventing the UV catastrophe. Alternatively, in certain systems, second quantization can be duly performed. However, the actual mechanism for preventing condensate formation (inverse cascade) or UV catastrophe (direct cascade) is not considered in this study as it only pertains to the inertial range. 
 
The absolute equilibrium distribution function is constructed as follows:
\begin{equation}
     \mathscr{P} = Z^{-1}\exp\lbrack-\alpha H -\beta H_M - \gamma H_C\rbrack =:
	Z^{-1}\exp\lbrack - A_{i,j} u^i u^j /2\rbrack,
	\label{partition}
\end{equation}
with $Z$ being the partition function. We observe that the above distribution function is divergent in the limit of large $k$. But, as we have remarked above, the dissipation range sets a natural cutoff for $k$, and thereby ensures that the distribution function does not blow up, since $k$ is bounded. This divergence arises because of the fact that $H$ is not in a `coercive' form - a similar feature was pointed out for Hall MHD in \citet{YoMa02}. Some of the above issues have been addressed successfully in the MHD context by \citet{ItYo96} and \citet{JYI98}. Given that we have considered only the inertial range, with a cutoff on $k$, the following discussion does not necessarily encompass absolute equilibria for the complete phase space.  

In the second equality of (\ref{partition}), note that the vector ${\bf u}$ is chosen to consist of 8 entries corresponding to 4 components (two real, two complex) of $\boldsymbol{v}_{\boldsymbol{k}}$ and $\boldsymbol{b}^*_{\boldsymbol{k}}$. We shall comment on the parameters $\alpha$, $\beta$ and $\gamma$ at a later stage in our discussion. The reduction in the total number degrees of freedom is due to solenoidal property of both fields: $\boldsymbol{k}\cdot\boldsymbol{b}^*_{\boldsymbol{k}}
	= 0 = \boldsymbol{k}\cdot\boldsymbol{v}_{\boldsymbol{k}}$.  Using \eqref{partition} we calculate the average of a quantity $F$ according to 
	\begin{equation}
	\langle F \rangle =\int \prod_{\bs{k}} d \bs{v}_{\bs{k}}\, d\overline{\bs{v}}_{\bs{k}}\,d\bs{b}^*_{\bs{k}}\, 
	d\overline{\bs{b}^*}_{\bs{k}} \; F\,  \mathscr{P}\,, 
	\label{average}
	\end{equation}
which will be used for all averages in the present section. Because all invariants are quadratic in $\bs{u}$ the integrations of \eqref{average} are all  Gaussian, allowing us to achieve our 	
goal of finding  correlations of the form $\langle u_i u_j\rangle = A^{-1}_{i,j}$; however, this   requires the inversion of the 8 by 8 matrix 
	\begin{equation}
		A=\begin{pmatrix}
			a & 0 & 0 & f \ 	&c & 0 & 0 & 0 \\
			0 & a & -f & 0 \ 	& 0 & c & 0 & 0 \\
			0 & -f & a & 0 \ 	& 0 & 0 & c & 0\\
			f & 0 & 0 & a \ 	&0 & 0 & 0 & c \\
			
			c & 0 & 0 & 0\ 				 	&d & 0 & 0 & b \\
			0 & c & 0 & 0 \ 					& 0 & d & -b & 0 \\
			0 & 0 & c & 0 \					& 0 & -b & d & 0 \\
			0 & 0 & 0 & c \ 					&b & 0 & 0 & d
		\end{pmatrix}\,,
	\end{equation}
	where $a:= \alpha$, $b = \beta/k$, $c = \gamma$, $f:=k(\beta d_e^2 + \gamma
	d_i)$ and $d := \alpha/(1+k^2 d_e^2)$. The inverse matrix fortunately has the same form as the simpler MHD case, and is given by
	\begin{equation}
		A^{-1}=\frac{1}{\Delta}\begin{pmatrix}
			P & 0 & 0 & X 	&Q & 0 & 0 & Y \\
			0 & P & -X & 0 	& 0 & Q & -Y & 0 \\
			0 & -X & P & 0 	& 0 & -Y & Q & 0\\
			X & 0 & 0 & P 	&Y & 0 & 0 & Q \\
			
			Q & 0 & 0 & Y				 	&R & 0 & 0 & W \\
			0 & Q & -Y & 0 					& 0 & R & -W & 0 \\
			0 & -Y & Q & 0 					& 0 & -W & R & 0 \\
			Y & 0 & 0 & Q 					&W & 0 & 0 & R
		\end{pmatrix}\,,
	\end{equation}
	where the new coefficients are
	\begin{equation}
		P:=	a(d^2-b^2)-c^2 d\quad\text{and}\quad X:=f(b^2-d^2)-c^2 b,
	\end{equation}
	\begin{equation}
		Q:= c(c^2 - a d - b f)\quad\text{and}\quad Y:= c(a b + d f),
	\end{equation}
	\begin{equation}
		R:= d(a^2-f^2)-c^2 a \quad\text{and}\quad W:= b(f^2-a^2)-c^2 f,
	\end{equation}
	\begin{equation}
		\sqrt{\det{A}} =: \Delta = (f b + a d - c^2)^2 - (a b + f d)^2.
	\end{equation}
The matrix $A$ has to be positive definite for the procedure to work, i.e.,  all of the eigenvalues must be positive \citep{frisch75}. The corresponding identities can be rearranged after a fair amount of algebra to arrive at the final set of positivity conditions
	\begin{equation}\label{inequality1}
		a>|f|,\quad d > |b|\quad\text{and}\quad c^2 < (a-|f|)(d-|b|).
	\end{equation}
A less strict, albeit useful, set of conditions  can be derived as well:
	\begin{equation}
		a d + b f > c^2\quad\text{and}\quad |a f + d b|<(a d
		+ b f - c^2)\quad\text{and}\quad |c| < \frac{a + d}{2}.
	\end{equation}
From these inequalities, we see that $\Delta > 0, P > 0$ and $R > 0$ as expected, ensuring that the autocorrelations are positive. Because of the normalized Alfv\'{e}n  scaling,  it is clear that $k>1$ must be valid, as otherwise we are concerning ourselves with length scales greater than the size of the system. Finally, the spectral quantities can be duly evaluated.

We write the Hamiltonian as the sum of kinetic and magnetic energies, $H=H_K +H_B$, with  the spectra of each given, respectively, by 
	\begin{eqnarray}
		E_K   &=& 2\pi k^2\sum_l \big \langle v_{l\boldsymbol{k}}\,  \overline{v}_{l\boldsymbol{k}}
		\big\rangle =
		\frac{8\pi k^2 P}{\Delta},
		\label{HKav}
\\
		E_B  &=& \frac{2\pi k^2}{1+k^2 d_e^2}\sum_l  \big \langle b^*_{l\boldsymbol{k}}\, 
		\overline{b^*}_{l\boldsymbol{k}} \big \rangle =
		\frac{8\pi k^2}{1+k^2 d_e^2}\frac{R}{\Delta}.
		\label{HBav}
	\end{eqnarray}
Similarly, the spectra of the  generalized magnetic and cross helicities, respectively,  are 
	\begin{eqnarray}
		E_M &=& 2\pi k^2 \sum_{l,m,n}\epsilon_{lmn}k_l
		\bigg(
		d_e^2 \big\langle v_{m\boldsymbol{k}}\, \overline{v}_{n\boldsymbol{k}} \big\rangle 
		+
		\frac{ 
		\big\langle b^*_{m\boldsymbol{k}} \, 
		\overline{b^*}_{n\boldsymbol{k}} \big\rangle}{k^2}
		\bigg)=
		8\pi k \, \frac{d_e^2 k^2 X + W}{\Delta},
\\
		E_C  &=& 
		2\pi k^2\sum_l\Big(2\big \langle v_{l\boldsymbol{k}}\, \overline{b^*}_{l\boldsymbol{k}}
		\big\rangle
		+\sum_{m,n} d_i\,\epsilon_{lmn}k_l \big \langle
		v_{m\boldsymbol{k}}\, \overline{v}_{n\boldsymbol{k}}\big  \rangle\Big) = 8\pi k^2\, 
		\frac{2 Q + d_i k X}{\Delta}.
	\end{eqnarray}
It is easy to obtain the spectra of  the original generalized helicities via the relation
$\mathcal{K}_\pm = 2(\kappa_\pm H_C + H_M)$, i.e.,  by  
\begin{equation}
K_{\pm}:=2\big(\kappa_\pm E_C + E_M\big)\,.
\label{Ks}
\end{equation}.

	
\subsection{Hall MHD Cascades}
\label{HMHDCas}

If we consider the Hall MHD limit as $1 < k \ll d_e^{-1}$, i.e. the range where Hall effects are important, we obtain the following conditions
	\begin{equation}\label{normalineq}
		\alpha> k|\gamma| d_i \quad\text{and}\quad \alpha > \frac{|\beta|}{k}
		\quad\text{and}\quad \gamma^2 < (\alpha-k|\gamma| d_i)\Big(\alpha-
		\frac{|\beta|}{k}\Big).
	\end{equation}
In addition, we also have
	\begin{equation}\label{normalineq2}
		\alpha^2 + \beta\gamma d_i >
		\gamma^2\quad\text{and}\quad \alpha > |\gamma|\quad\text{and}\quad \alpha^2 >
		|\beta\gamma| d_i.
	\end{equation}
During the process of computing the last inequality in \eqref{normalineq} for $k$, we also computed the discriminant
	\begin{equation}\label{discriminant}
		\mathcal{D}:= (\alpha^2 + |\gamma\beta| d_i - \gamma^2)^2 - 4
		\alpha^2|\gamma\beta| d_i.
	\end{equation}
Requiring the existence of a $k$-spectrum ($\mathcal{D}>0$) leads us to a stricter version of the first inequality in \eqref{normalineq2}:
	\begin{equation}\label{strictineq}
		\alpha > |\gamma| + \sqrt{|\gamma\beta| d_i}.
	\end{equation}

To see how this inequality is obtained, let us rewrite \eqref{discriminant} as
\begin{equation}
	0<\mathcal{D} = \big((\alpha - \sqrt{|\gamma\beta| d_i})^2 - \gamma^2\big) (\alpha^2 + |\gamma\beta| d_i - \gamma^2+2\alpha\sqrt{|\gamma\beta| d_i}).
\end{equation}
The second term in the product is clearly positive according to \eqref{normalineq2}. Thus, one requires the first term to be positive which leads us to \eqref{strictineq}. In turn, this leads us to stricter requirements on $k$ than the ones of the first two inequalities in \eqref{normalineq}. Our bounds are thus given by
	\begin{equation}
		\frac{|\beta|}{\alpha}<\frac{\alpha^2 + |\beta\gamma| d_i - \gamma^2 -
		\sqrt{\mathcal{D}}}{2\alpha|\gamma| d_i} < k < \frac{\alpha^2 + |\beta\gamma|
		d_i - \gamma^2 + \sqrt{\mathcal{D}}}{2\alpha|\gamma| d_i} <
		\frac{\alpha}{|\gamma| d_i}.
	\end{equation}
The lower bound on $k$ is also present in ideal MHD, but the upper limit appears to be solely due to the inclusion of the Hall term. Notice that if we wish to extend the range of $k$ much further beyond $d_i^{-1}$ it is reasonable to impose $\alpha\gg|\gamma|$. Therefore, since $d_i\ll 1$ is typically valid, the assumption  $\alpha^2-\gamma^2\gg |\gamma\beta| d_i$ is also justified. If we use this, along with an expansion in $d_i$, the limits can be approximated as
	\begin{equation} \label{ApproxInd}
		\frac{\alpha|\beta|}{\alpha^2-\gamma^2} \lesssim k \lesssim
		\frac{\alpha^2-\gamma^2}{\alpha|\gamma| d_i}
	\end{equation}
so that the parameters can be adjusted to allow for $1 < k \ll d_e^{-1}$. 

In the Hall limit the different spectral densities are given by
	\begin{eqnarray} \label{EKinH}
		E_K &=& 8\pi\alpha\frac{k^2(\alpha^2-\gamma^2) -
		\beta^2}{(\alpha^2+\beta\gamma d_i-\gamma^2)^2 - \alpha^2(k\gamma d_i
		+\beta/k)^2},
\\ \label{EMagH}
		E_B  &=& 8\pi\alpha k^2\frac{\alpha^2-\gamma^2 -
		\gamma^2 k^2 d_i^2}{(\alpha^2+\beta\gamma d_i-\gamma^2)^2 - \alpha^2(k\gamma
		d_i +\beta/k)^2},
\\
		 E_M &=&  8\pi\frac{\gamma^2 d_i(\beta
		d_i-\gamma)k^2-\beta \alpha^2}{(\alpha^2+\beta\gamma d_i-\gamma^2)^2 - \alpha^2(k\gamma d_i
		+\beta/k)^2},
\\
	  E_C &=&  8\pi\gamma
		k^2\frac{d_i^2(\beta^2-\alpha^2 k^2)-\gamma\beta d_i-2(\alpha^2+\beta\gamma
		d_i - \gamma^2)}{(\alpha^2+\beta\gamma d_i-\gamma^2)^2 - \alpha^2(k\gamma d_i +\beta/k)^2}.
	\end{eqnarray}
We note that each of these spectra are identical to the previous expressions obtained by \citet{SMC08} (see their Eqs.~(26)-(29)), after undertaking a minor change of variables. This is not surprising as the authors had derived them using the same approach, viz. by constructing the absolute equilibrium states. We also wish to point out an important result that has also been predicted by many others before - the absence of equipartition between the kinetic and magnetic spectra in Hall MHD \citep{GSRG96,KM04,Gal06,GB07,SMC08,SS09,LinB16,LB16}. This trait is  unique to Hall MHD, as it is absent both in ideal MHD and inertial MHD; we shall demonstrate the latter in Sec.~\ref{IMHDCas}.	
	
Notice that the average  total energy spectrum $E=E_K + E_B$  can be computed from (\ref{EKinH}) and (\ref{EMagH}), and has the form
\begin{equation}
		E  = 8\pi\alpha k^2\frac{2(\alpha^2+\beta\gamma d_i-\gamma^2) -
		(k\gamma d_i +\beta/k)^2}{(\alpha^2+\beta\gamma
		d_i-\gamma^2)^2 - \alpha^2(k\gamma d_i +\beta/k)^2},
\end{equation}
which is also equal to the formula provided in \citet{SMC08}.
The parameters $\alpha$, $\beta$,  and $\gamma$ are found by matching the integrated spectral quantities with their actual spatial values,  e.g.,   by using  $\int_{k_{min}}^{k_{max}} E_K  \,  dk$,
where we imagine a continuum limit. Thus, it is obvious that one cannot provide simple expressions for these parameters, since they will be complicated transcendental equations in general. 

As noted earlier, the dependence of the spectral quantities on $k$ will reveal the directionality of the cascades. The	direct cascade of some invariant can be expected if the spectral density is peaked at high wavenumbers and vice-versa. Based on the complexity of the above formulae even for Hall MHD, it appears as though any definitive statements are not possible. It is reasonable to expect that the same quantity may undergo both cascades depending on the length scale at which the energy is supplied to the system.
	
The simplest case one can investigate is to consider cases where the cross-helicity vanishes, viz.\  $E_C = 0\Rightarrow \gamma = 0$.  For this case we have verified that the standard MHD results presented in 
\citet{frisch75} are obtained, i.e.,  the direct cascade of energy and the inverse cascade of magnetic helicity. This result is not  at all surprising  because  the magnetic helicity is an invariant of both  ideal and Hall MHD. Our analysis confirms that, in the absence of global cross-helicity, when magnetic helicity is injected at length scales much larger than the electron skin depth, it undergoes an inverse cascade within the framework of Hall MHD.
	
Let us consider another simple limit, where $\beta = 0$. At first glimpse, it doesn't have such a simple interpretation. We can make the picture more transparent by introducing the definitions
	\begin{equation}
		\frac{\gamma}{\alpha}=: \sin{\phi}\quad\text{and}\quad
		\frac{\cos^2{\phi}}{|\sin{\phi}|\, d_i}  =: k_* > k,
	\end{equation}
where  the second equality follows from the second relation in (\ref{ApproxInd}). The corresponding spectral quantities in these new variables are thus given by
	\begin{equation}
		E_K  = \frac{8\pi}{\alpha \cos^2{\phi}}
		\frac{k^2}{1-\dfrac{k^2}{k_*^2}}\quad\text{and}\quad E_B  =
		E_K \Big(1-\frac{k^2}{k_*^2} \cos^2{\phi}\Big),
	\end{equation}
	together with
	\begin{equation}
		E_M  = - d_i E_K \sin{\phi} \tan^2{\phi}\quad\text{and}\quad
		 E_C = -E_K 
		\sin{\phi}\Big(2+\frac{k^2}{k_*^2}\cot^2{\phi}\Big).
	\end{equation}
After a careful inspection and evaluation, one can verify that these expressions yield direct cascades of energy and cross-helicity. 
	
\begin{figure}[h]
		\begin{center}
			\subfloat[Total Energy]{
				\includegraphics[width=.33\textwidth]{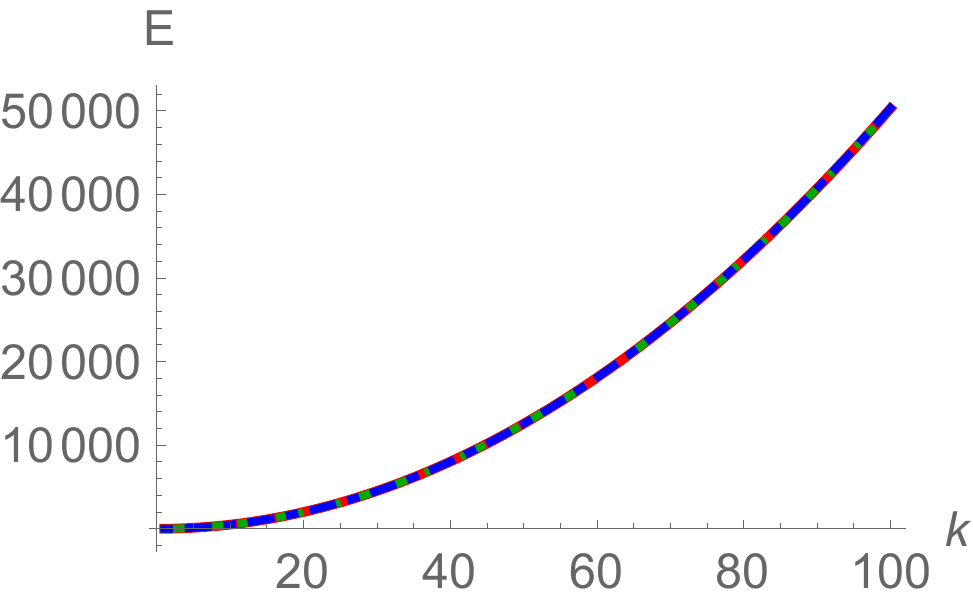}
			}
		\subfloat[Cross Helicity]{\includegraphics[width=.33\textwidth]{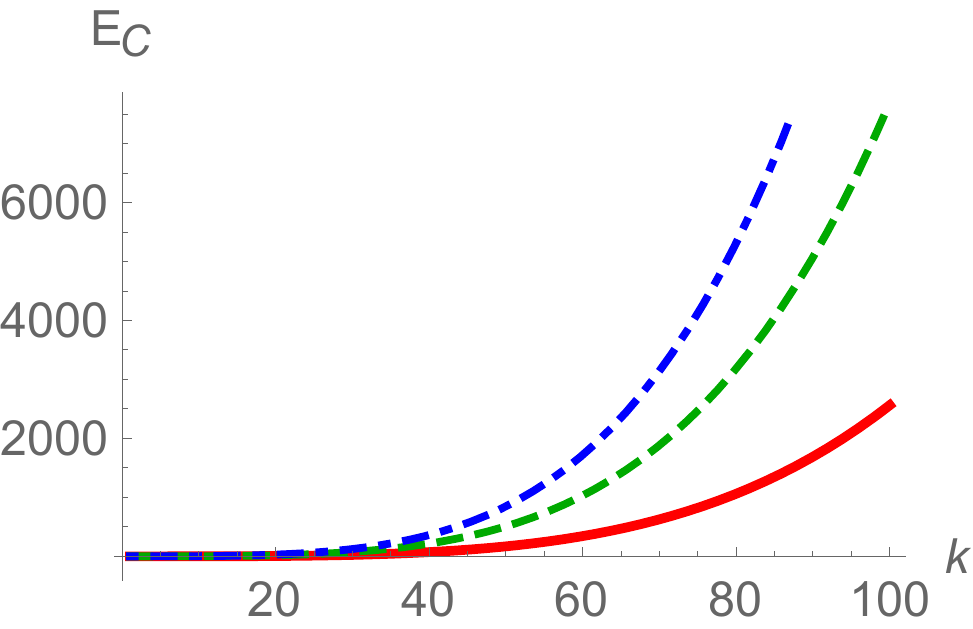}
		}
		\subfloat[Magnetic Helicity]{\includegraphics[width=.33\textwidth]{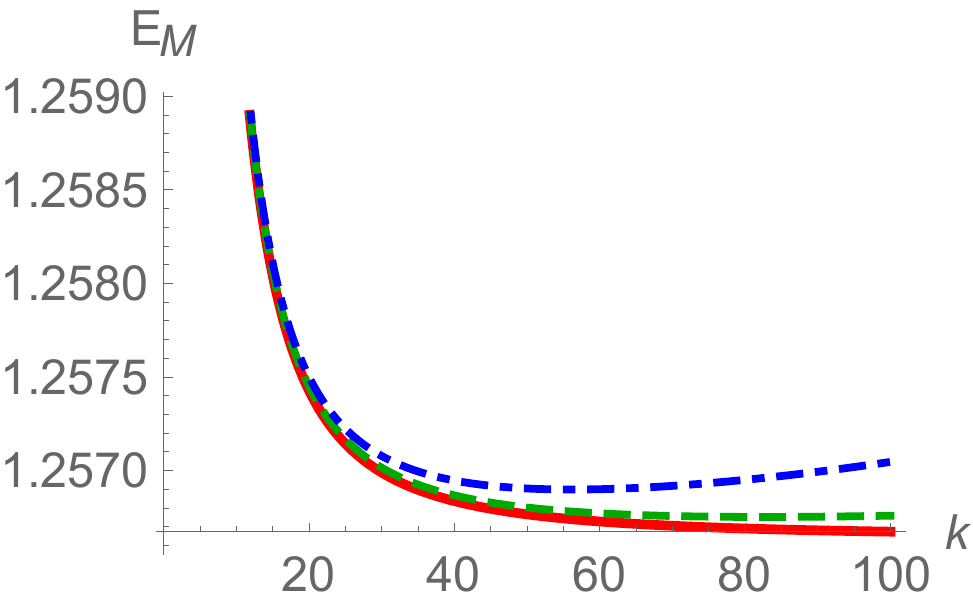}
		}
		\end{center}
	\caption{\small Plots of absolute equilibrium  states for absolute values of spectral quantities for  Hall MHD. The parameters chosen are $\alpha = 10$, $\beta = 5$, $d_i = 0.1$. The third parameter is varied with  the solid red line corresponding  to $\gamma=0.01$ with $\langle H_C\rangle/\langle H \rangle\approx 0.03$, the dashed green line corresponding  to $\gamma = 0.03$ with $ \langle H_C \rangle/ \langle H \rangle\approx 0.09$, and the dot-dashed blue line corresponding  to $\gamma = 0.05$ with  $\langle H_C \rangle/ \langle H \rangle\approx 0.16$. The spectral range is chosen to be $1<k<d_i^{-2}$.
	}\label{hallcascade}
	\end{figure}

In order to visualize these relations, we have plotted the different spectra in Fig.~\ref{hallcascade}. It is particularly noteworthy that the magnetic helicity cascade becomes increasingly complex in the presence of strong cross-helicity. This is purely due to the additional perturbation coming from the Hall term, as the ideal MHD range remains completely in the inverse cascade mode.
	
Hence, we sum up this preliminary analysis by observing that $\mathcal{K}_{+}$ can undergo both forward and inverse cascades as predicted by \citet{BG16}. We have also verified that our XMHD spectra, in the Hall MHD limit, are equal to the ones obtained earlier by \citet{SMC08}.

	
\subsection{Inertial MHD Cascades}
 \label{IMHDCas}

We begin by recalling that inertial MHD is a model which lacks the Hall drift, but is endowed with electron inertia effects \citep{KM14,LMT14}.	Thus, the existence of the second condition implies that the model may become relevant in the range $k\gg d_e^{-1}$, i.e. at scales smaller than electron skin depth. Although this quantity is small in many fusion plasmas, recall that it is highly relevant in astrophysical and space plasmas, such as the Earth's magnetosphere and the solar wind. With this choice of $k$, observe that $d \approx \alpha/k^2 d_e^2$ holds true. 

Following the same procedure as in Hall MHD, we analyze the necessary inequalities, and find that
	\begin{equation}
		\alpha>k|\beta|d_e^2\quad\text{and}\quad \alpha \gtrsim |\gamma|.
	\end{equation}
Although $d_e \ll 1$, we also have $k d_e \gg 1$ in this case, and hence the condition $\alpha\gtrsim |\beta|$ appears to be quite reasonable. We also must inspect a counterpart of the 
third inequality in \eqref{inequality1}, which according to the constraints listed above collapses to
	\begin{equation}
		|\gamma| < \frac{\alpha}{k d_e}-|\beta|d_e.
	\end{equation}
Upon computation, the spectra become
	\begin{eqnarray}
	E_K   &=& E_B   = \frac{8\pi\alpha}{d_e^2}\frac{\dfrac{\alpha^2}{k^2 d_e^2}-\beta^2 d_e^2-\gamma^2}{\Big(\beta^2+\dfrac{\alpha^2}{k^2 d_e^2}-\gamma^2\Big)^2-\dfrac{4\alpha^2}{\beta^2 k^2}},
\\
	E_M  &=& -16\pi \beta k^2 d_e^2\frac{\dfrac{\alpha^2}{k^2 d_e^2}+\gamma^2-\beta^2 d_e^2}{\Big(\beta^2+\dfrac{\alpha^2}{k^2 d_e^2}-\gamma^2\Big)^2-\dfrac{4\alpha^2}{\beta^2 k^2}},
\\
	E_C  &=& -16\pi \gamma k^2 \frac{\beta^2 d_e^2+\dfrac{\alpha^2}{k^2 d_e^2}-\gamma^2}{\Big(\beta^2+\dfrac{\alpha^2}{k^2 d_e^2}-\gamma^2\Big)^2-\dfrac{4\alpha^2}{\beta^2 k^2}}.
	\end{eqnarray}
An important and pleasing feature is immediately apparent. We see that inertial MHD restores the energy equipartition feature of ideal MHD \citep{Shiv15}. This is along expected lines,  since inertial MHD and ideal MHD are very akin to each other. In fact, it was shown by \citet{LMT14} in 2D that the Hamiltonian (Poisson bracket) structure of these two models is identical under the transformation ${\bf B} \rightarrow {\bf B}^*$. 
	
We also see that the generalized magnetic and cross helicities vanish when $\beta$ and $\gamma$ are set to zero respectively. Hence, it is instructive to take these two limits and inspect the resultant expressions. When $\beta = 0$, the total energy is
	\begin{equation}
	E = \dfrac{16\pi\alpha}{\dfrac{\alpha^2}{k^2}-\gamma^2 d_e^2},
	\end{equation}
	and the cross-helicity is
	\begin{equation}
		E_C  = -  E   \frac{\gamma}{\alpha} k^2 d_e^2.
	\end{equation}
	
\begin{figure}[h]
		\begin{center}
			\subfloat[Total Energy]{
				\includegraphics[width=.33\textwidth]{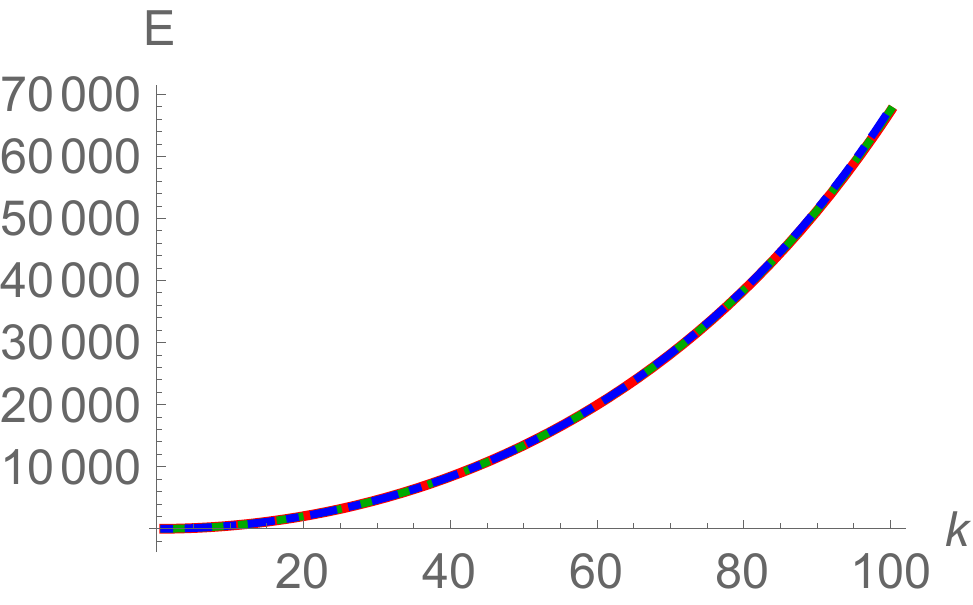}
			}
			\subfloat[Cross Helicity]{\includegraphics[width=.33\textwidth]{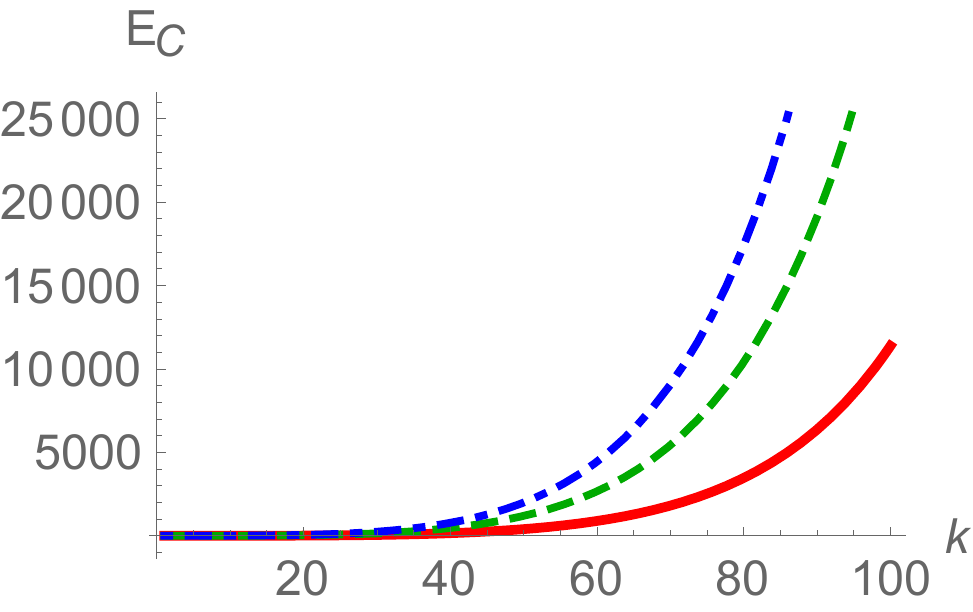}
			}
			\subfloat[Magnetic Helicity]{\includegraphics[width=.33\textwidth]{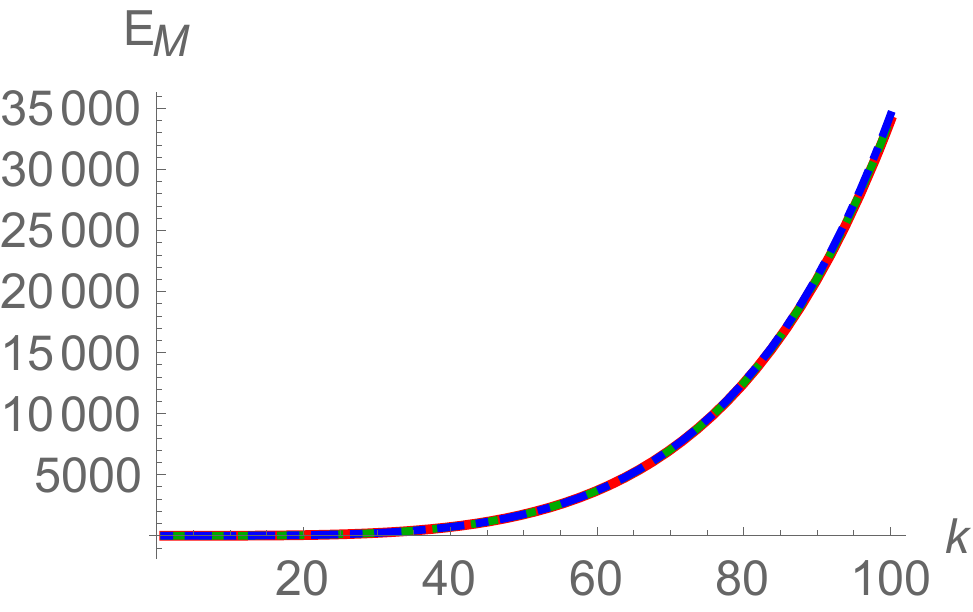}
			}
		\end{center}
		\caption{\small Plots for absolute equilibria states  of spectral quantities for  inertial MHD. The parameters chosen are $\alpha = 10$, $\beta = 5$, $d_e = 0.1$. The third parameter is varied with  the solid red line corresponding  to $\gamma=0.01$ with $\langle H_C \rangle/  \langle H \rangle \approx 0.09$, the dashed green line to $\gamma = 0.03$ with $\langle H_C \rangle/  \langle H \rangle \approx 0.28$ and the dot-dashed blue line to $\gamma = 0.05$ with $\langle H_C \rangle / \langle H \rangle \approx 0.46$. The spectral range is chosen to be $1<k<d_e^{-2}$. 
		}\label{inertialcascade}
	\end{figure}

The other case, with $\gamma = 0$, corresponds to the state with zero cross-helicity. In this instance, we find that the spectra are
	\begin{equation}
		E    = \dfrac{16\pi\alpha}{\dfrac{\alpha^2}{k^2}-\beta^2 d_e^4}
		\quad\text{and}\quad E_M    = - E    \frac{\beta}{\alpha} k^2 d_e^4.
	\end{equation}
In each of these two limiting cases, we find that all spectral quantities undergo direct cascades in contrast to the MHD and Hall MHD limits. This appears to be consistent, to an extent, with previous results in the literature although most previous  studies relied on 2D simulations as opposed to our 3D analysis \citep{BSD96,BSZCD,DDK00}. We have plotted the spectra in Fig.~\ref{inertialcascade},  which confirms our theoretical predictions. Although the wavenumber range from $k$ to $1/d_e^2$, for  inertial MHD is not applicable everywhere - instead, its presence is likely to be felt only when $k > 1/d_e$. In reality, there is a finite Hall MHD range before this limit is attained.

We note, in passing, that the inclusion of a strong guide field can induce  anisotropic turbulence, and the existence of both inverse and direct cascades, but this falls outside the scope of our present work.
	

	\section{Discussion and Conclusion} \label{SecConc}

There has been a great deal of attention in recent times focused on turbulence at `small' scales, i.e., scales  smaller than the electron or proton gyroradius (or skin depth). The two most notable examples in astrophysics are   the Earth's magnetosphere and the solar wind,  respectively. The recently launched \emph{Magnetospheric Multiscale} (MMS) Mission is known to be capable of probing such scales \citep{Butal16}, and observational results in these regimes have also been recently published \citep{Nar16}. We also note that probing such scales may also become  feasible in the laboratory, such as the WiPAL \citep{FF15}.

Thus, in the coming years, it is likely that a thorough understanding of the physics at these scales will be necessary. To gain such an understanding, it is imperative to work with models that are applicable at such small scales. Neither ideal nor Hall MHD are valid when one approaches length scales on the order of the electron skin depth. As extended MHD (XMHD) is endowed with both the Hall drift and electron inertia, it constitutes a good physical model in this regime. Of course, we must caution the reader that it does not capture certain kinetic effects such as Landau damping, pressure anisotropy, etc. and also lacks dissipative effects. 

In this paper, we have first focused on gaining a basic understanding of the salient mathematical properties of XMHD. This was done by taking recourse to the Hamiltonian formulation, which is endowed with several advantages. In particular, we discussed how the mutually exclusive effects of electron inertia and the Hall current can be unified into a single framework. We also presented the helicities of XMHD that are generalizations of the magnetic/fluid helicity. This was done by demonstrating that they are topological invariants of XMHD and can be determined through the Hamiltonian approach. 

All of these facts were duly invoked in the subsequent sections, where we focused on some aspects on XMHD turbulence. Firstly, we generalized the results of \citet{BG16}, where the dissipation rates of Hall MHD were computed by using second order structure functions. We showed that our results, with electron inertia, still resembled the Hall MHD case, and reduced to the latter when the electron skin depth was vanishingly small. We also showed that, in the limit of infinite Reynolds number, the dissipation rates vanished when the Beltrami conditions were satisfied, thereby confirming earlier predictions. 

	\begin{table*}[htpb]
		\caption{\small Direction of cascades in MHD, HMHD and IMHD. *When $\langle H_C \rangle \ll \langle H \rangle $}
		\vspace{.25cm}
		\label{t:induction equations}
		\centering
		{\renewcommand{\arraystretch}{1.5}%
			{\setlength{\tabcolsep}{1.0em}
				\begin{tabular}{|r|c|c|c|}
					\hline
					& Ideal MHD & Hall MHD & Inertial MHD \\
					\hline
					$E$ & direct  & direct & direct \\
					\hline
					$E_M$ & inverse & inverse* & direct \\
					\hline
					$E_C$ & direct & direct & direct \\
					\hline
					${K}_-$ & inverse  & inverse*  & direct \\
					\hline
					${K}_+$ & inverse & both & direct \\
					\hline
				\end{tabular}
			}
		}
	\end{table*}

The central theme of the paper, however, revolved around the issue of the directionality of the cascades of the helicities and the energy. Unlike ideal MHD, where a direct cascade of the energy and the inverse cascade of the magnetic helicity can be unambiguously predicted, the situation is rendered far more complex due to the Hall drift and electron inertia. We commenced our analysis by proving Liouville's theorem for the first time for  XMHD, which was necessary for constructing the absolute equilibrium states of XMHD. The latter were used to study the cascades of XMHD in two limiting regimes: (i) where the Hall term is important and the electron inertia terms   unimportant, and (ii) vice-versa. 

In the Hall regime, the energy and the magnetic helicity still exhibit direct and inverse cascading, respectively. However, the ion canonical helicity $\int_D d^3x\,\left({\bf A} + d_i {\bf V}\right) \cdot \left({\bf B} + d_i \nabla \times {\bf V}\right)$, which is a conserved quantity in Hall MHD, can undergo a cascade in either direction. We also verified that the magnetic and kinetic energy spectra are characterized by a lack of equipartition, which constitutes a staple and unique feature of Hall MHD (that is absent in ideal MHD). Each of these results were shown to be consistent with, or identical to, previous studies; see, for e.g. \citet{SMC08}. 

When electron inertia effects were taken to be dominant over the Hall term (the inertial MHD regime) we found that equipartition was recovered. In addition, we also found that all of the quantities, viz.\  the energy and the two helicities, undergo direct cascading in this regime. Hence, we expect that the (generalized) magnetic helicity undergoes inverse cascading  up to a certain length scale (for a given choice of the free parameters), and then undergoes a reversal, consequently ending up as a direct cascade. A summary of these results can be found in Table \ref{t:induction equations}.

In addition to the aforementioned systems such as the solar wind and Earth's magnetosphere, our results may also have significant consequences in other areas. The presence of the inverse cascade of magnetic helicity is intimately linked to the generation of magnetic fields via the dynamo mechanism \citep{Bran01,BS05,AMP06}. The existence of an inverse cascade has already been investigated in the Hall regime by   \citet{PDM03,MAP07,LB16}. But, if this feature were to be non-operational at smaller scales, it may lead to non-trivial, potentially far reaching, consequences in dynamo theory. Lastly, our analysis is also likely to be of some relevance in turbulent reconnection, which remains an active and unresolved area of research \citep{Serv11,LEV12}. Hence, on account of the aforementioned reasons, we suggest further analyses of this kind are timely and warranted.

In closing, we describe a couple of important avenues that can be explored by means of our formalism.  To begin with, the 2D version (reduced XMHD) can also be subjected to the same treatment since its Hamiltonian structure and invariants have been thoroughly studied and classified \citep{GTAM16}. Alternatively, electron-positron plasmas, which are produced both via high-intensity lasers in the laboratory \citep{MTaB06}, and also occur in many astrophysical settings \citep{HL06}, have attracted a great deal of interest recently. The dynamical equations for these plasmas are characterized by the absence of the Hall term, and can be derived by adopting a procedure akin to \citet{Dung58} and \citet{Lust59}. It is, therefore, advantageous to utilize the Hamiltonian structure for this model \citep{KMM16}, to derive the corresponding invariants (such as the generalized helicities), and to carry out an analysis of the cascades and spectra along the lines of our present work.

In recent times, there has been a great deal of interest on  the study of relativistic turbulence, given its importance in laboratory and astrophysical plasmas. Most of the works thus far have focused on computational or phenomenological studies, as seen from the likes of
\citet{KR09,ZY11,IAI11,Zra14}. In contrast, it has recently been shown that relativistic MHD possesses a noncanonical Hamiltonian formulation \citep{DAMP15}. Thus, it is evident that a study akin to the one presented herein could be undertaken for relativistic MHD as well. We conclude by observing that one can combine relativity and two-fluid effects to arrive at relativistic extended MHD, which is also known to have a Hamiltonian structure \citep{KMM16}. On account of the model's generality, we suggest that this could be used for constructing the spectra and cascades, and most of the results (and models) discussed here would automatically follow as limiting cases.

\acknowledgments
\noindent ML was supported by the NSF (Grant No. AGS-1338944) and the DOE (Grant No. DE-AC02-09CH-11466) during the course of this work. PJM and GM received support from the DOE  (Grant No.  DE-FG05-80ET-53088). PJM would also like to acknowledge support via a  Forschungspreis from the Humboldt Foundation and the hospitality of the Numerical Plasma Physics Division of IPP, Max Planck, Garching.

\bibliographystyle{apsrev4-1}
\bibliography{xcascades}

\end{document}